# DEVICE FOR MEASUREMENT OF THERMAL EMISSIVITY AT CRYOGENIC TEMPERATURES

Tomas Kralik, Pavel Hanzelka, Vera Musilova, Ales Srnka

Institute of Scientific Instruments ASCR, Kralovopolska 147, Brno, CZ – 612 64, Czech Republic
email: kralik@isibrno.cz

## ABSTRACT

In the described device, the thermal emissivity or absorptivity of the sample is measured by substitution of the radiative heat flow between two parallel surfaces by thermal output of a heater. Fast measurements of the mutual emissivity for the range of the temperature of the radiating surface 25 K-150 K are possible. The absorbing surface has a temperature between 5 K and 10 K when LHe is used as cryoliquid. The desired measurement sensitivity is 1 mK for temperature and 0.1 μW for heat power, respectively. The diameter of the whole device is 50 mm and so it is possible to use a commercial dewar can for the cooling. The form of the sample is a round plate 40 mm in diameter and 1 mm in thickness with one tested side. The emissivity and its temperature dependency for various surface treatments can be checked immediately before application in a cryogenic system.

## 1. INTRODUCTION

The knowledge of thermal radiative properties of surfaces is important for design of low temperature apparatuses. Our Laboratory of cryogenics needs these values for designing and construction of various experimental cryostats and applications.

Existing published data are often not sufficiently useable. Most data are obtained at a sample temperature of 4.2 K, 77 K and 300 K and a temperature of radiation of 77 K and 300 K (Scott, 1959, Kaganer, 1969). Very little experimental data about temperature dependency exist. It is possible to found some spectral data (Biondi and Guobadia, 1968, Bos and Lynch, 1970, Tsujimoto *et al.*, 1982) for samples at 4 K.

## 2. THE MUTUAL EMISSIVITY

For the measurement of the radiative heat transfer between two parallel surfaces (radiator and absorber) under cryogenic conditions a new apparatus was developed at ISI Brno. It is possible to vary a temperature of the radiator $T_R$ and the absorber $T_A$. The measurement is based on the detection of the radiative heat flux $Q_R$ absorbed by the absorber and on calculation of the mutual emissivity $\varepsilon_{RA}$ by (1).

$$\varepsilon_{RA} = \frac{Q_R}{A\sigma(T_R^4 - T_A^4)} \qquad (1)$$

*A* is an area of the absorber (equal to the radiator area), σ is the Stefan-Boltzmann-constant.

If radiative properties do not depend on the wavelength of thermal radiation, the mutual emissivity of two parallel surfaces may be calculated (2) from the total hemispherical emisivity $\varepsilon(T_R)$ of the hot surface and total hemispherical absorptivity $\alpha(T_R, T_A)$ of the cold surface.

$$\frac{1}{\varepsilon_{RA}} = \frac{1}{\varepsilon} + \frac{1}{\alpha} - 1 \qquad (2)$$

If the emissivity $\varepsilon(T_R)$ of a radiator (e.g. metallic surface) is very small in comparison to the absorptivity $\alpha(T_R, T_A)$ of an absorber surface, the measured mutual emissivity is approximately equal to the emissivity of the radiator.

In such a configuration total hemispherical emissivity of metallic surface may be measured. And vice versa, when metallic sample is in the position of the absorber, its total hemispherical absorptivity is measured.

## 3. DESCRIPTION OF THE APPARATUS

The isothermal measuring chamber (IMC, Figure 1) is placed in the evacuated stainless steel casing tube (SCT) of a diameter of 48 mm. The length of the whole apparatus is about 1 m. This form allows the cooling by using of a commercial dewar with a neck diameter of 50 mm.

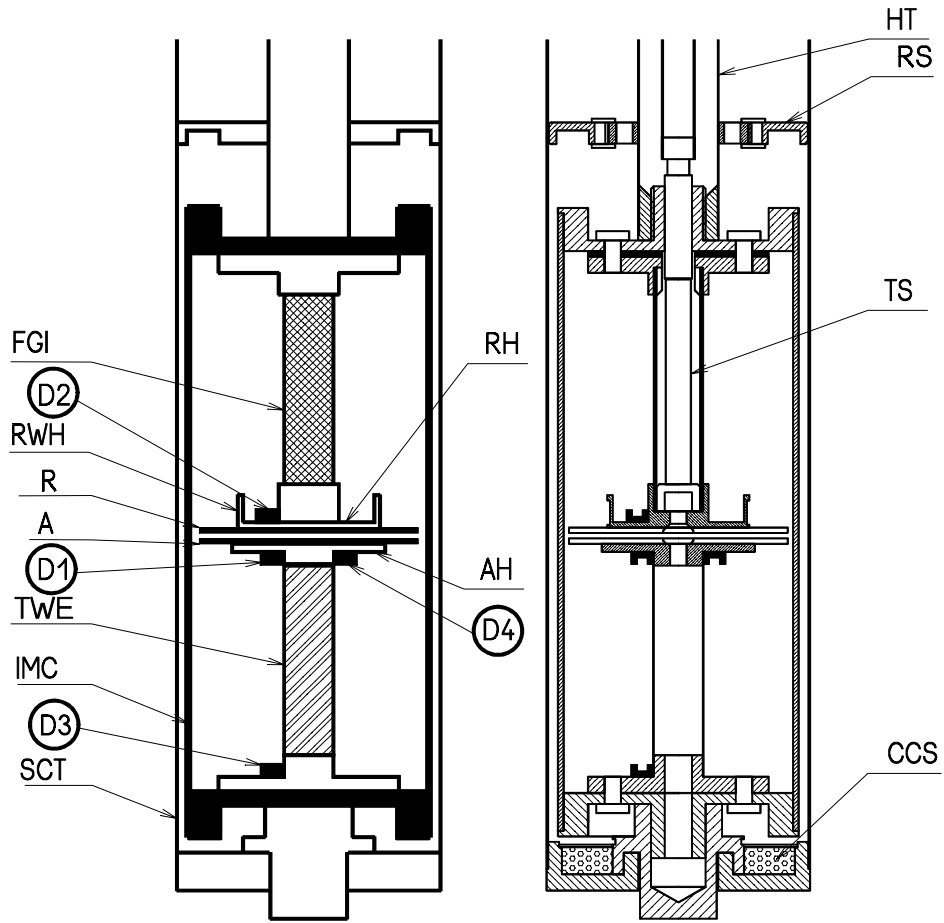

Figure 1 - Bottom part of the apparatus

> Left: D1,D2,D3 - diodes (Lake Shore DT–470), D4 - diode (Lake Shore DT-470) as microwatt heater $Q_{HA}$, A - absorber, AH – absorber holder, R-radiator, RH – radiator holder, RWH – resistive wire heater $Q_{HR}$, STR – stainless steel thermal resistor, FGI - fibre glass insulator, IMC – isothermal measuring chamber, SCT – evacuated stainless steel casing tube
> 
> Right: RS – radiation shield, TS – thermal switch, CCS – charcoal sorbent

The bottom of the casing tube (CT) is welded together with the stainless plug (SP). Copper thermal contact (CTC) with a thread is soldered in this plug. The thread thermally joins the measuring chamber with the thermal contact and so with the cryoliquid in the dewar. The plug contains a small chamber with charcoal sorbent (CCS). The whole measuring chamber is mounted to the holding tube (HT). Electrical contacts and wires are fixed to this tube. Four radiation shields (RS) attached to the holding tube support the thin casing tube against surrounding pressure. Each shield is cooled by friction contact with the casing tube. An actuating tube of thermal switch (TS) leads through the holding tube.

The upper end of the casing tube is terminated with a pot covered by an O-ring sealed plate. The plate is equipped with three welded vacuum connectors and a thermal switch actuator.

The radiator (R) and the absorber (A) (both diam. 40 mm, thick. 1 mm) are situated in the measuring chamber (IMC). Between the absorber and the radiator is a gap of about 0.5 mm.

The radiator is attached to the radiator holder (RH) heated by a resistance wire heater (RWH). The radiator holder is linked with the top of the measuring chamber by a thermal insulator (TIF) made of thin fibreglass tube.

The absorber is attached to the absorber holder (AH) heated by a microwatt calibration heater. Diode D4 Lake Shore DT–470 is used as this heater. The thermal contact is improved by Apiezon grease (Gmelin, 1999). The holder of the absorber is linked to the bottom of the measuring chamber by the thermal resistor (STR) made of a thin stainless steel tube.

The inner sides of the measuring chamber are coated with an epoxy layer. This "black" layer inhibits multiple reflections of thermal radiation coming from the backside of the radiator and prevents this parasitic heat from penetration into the gap between the radiator and absorber.

For the measurement of temperatures $T_A$ (absorber), $T_R$ (radiator) and $T_K$ (chamber bottom) three calibrated silicon diodes (D1, D2, D3) Lake Shore DT–470 are used.

## 4. MEASUREMENT PROCESS

Evacuated apparatus is firstly cooled down in $LN_2$ (30 min) and then it is inserted in the LHe dewar. The cool-down process of the whole apparatus in LHe takes about 8 hours and is carried out overnight. Slow heat transfer from the absorber through thermal resistor determines the time of cooling. The thermal switch accelerates cooling of the radiator. Constant temperature of LHe bath (Hanzelka and Musilova, 1995) is ensured by a pressure stabilizer (Hanzelka and Jura, 2002).

The main principle of the measurement method is substitution of the radiative heat flow $Q_R$ by the electric power $Q_{HA}$. For this purpose the calibration of thermal resistor is performed, i.e. the dependence $Q_{HA}(T_A-T_K)$ is found.

The radiator has the temperature of the chamber during calibration. The heat radiated from the backside of the absorber is negligible in comparison with conductive heat flow. Then the emisivity measurement follows.

The radiator temperature $T_R$ is adjusted by heating with $Q_{HR}$ and absorber temperature $T_A$ is measured after reaching its equilibrium value. Radiative heat flow $Q_R$ is calculated from the calibration $Q_{HA}(T_A-T_K)$ and gradient $T_A-T_K$. The relation (1) gives mutual emissivity value. If an elevated temperature of the absorber is required during the emissivity measurement, an additional heat flow produced by $Q_{HA}$ is applied.

The temperatures $T_A$ and $T_K$ are detected with a sensitivity of 1 mK. This sensitivity enables measurement of radiative heat flows as low as 0.1 µW.

Some amount of the radiative heat flow $Q_R$ leaks out from the gap between the absorber and radiator. Our measurements for the same sample and variable gap show that the error of the emissivity caused by this leak of the radiation is not greater than 10 % if the gap width is maximally 1 mm. This error is systematic and can be corrected. Usually a gap of 0.5 mm was adjusted.

## 5. EXPERIMENTAL RESULTS

At this time our configuration allows measurement of absorptivity of metals. A copper plate covered by epoxy layer 60 µm thick is used as the radiator. This layer does not represent a perfect "black body" surface. The mutual emissivity of two epoxy surfaces at $T_A$=5.5 K and $T_R$=30.4 K is 34 %. But this epoxy "blackness" insures that small values of mutual emissivity of the sample - epoxy configuration are with high accuracy identical to the value of sample absorptivity. For example, a mutual emissivity of 5% corresponds to a sample absorptivity of 5.05%.

For the samples with higher absorptivity the measured value of mutual emissivity must be recalculated.

### 5.1 TESTS OF THE METHOD

The method was tested by the following experiment. We measured absorptivity of a chemically polished copper first. Then we made an epoxy spot (1.4 % of the sample area, 70 μm thick) on the sample surface and the apparent absorptivity of the sample was measured again. The last step was the measurement of the mutual emissivity of two epoxy surfaces.

The apparent absorptivity of the surface with the spot was predicted by two models. The first one supposed the influence of the spot on the radiation density in the gap. The second one omitted this influence. The measured apparent absorptivity of the surface with the spot lies between predictions of both models (Figure 2).

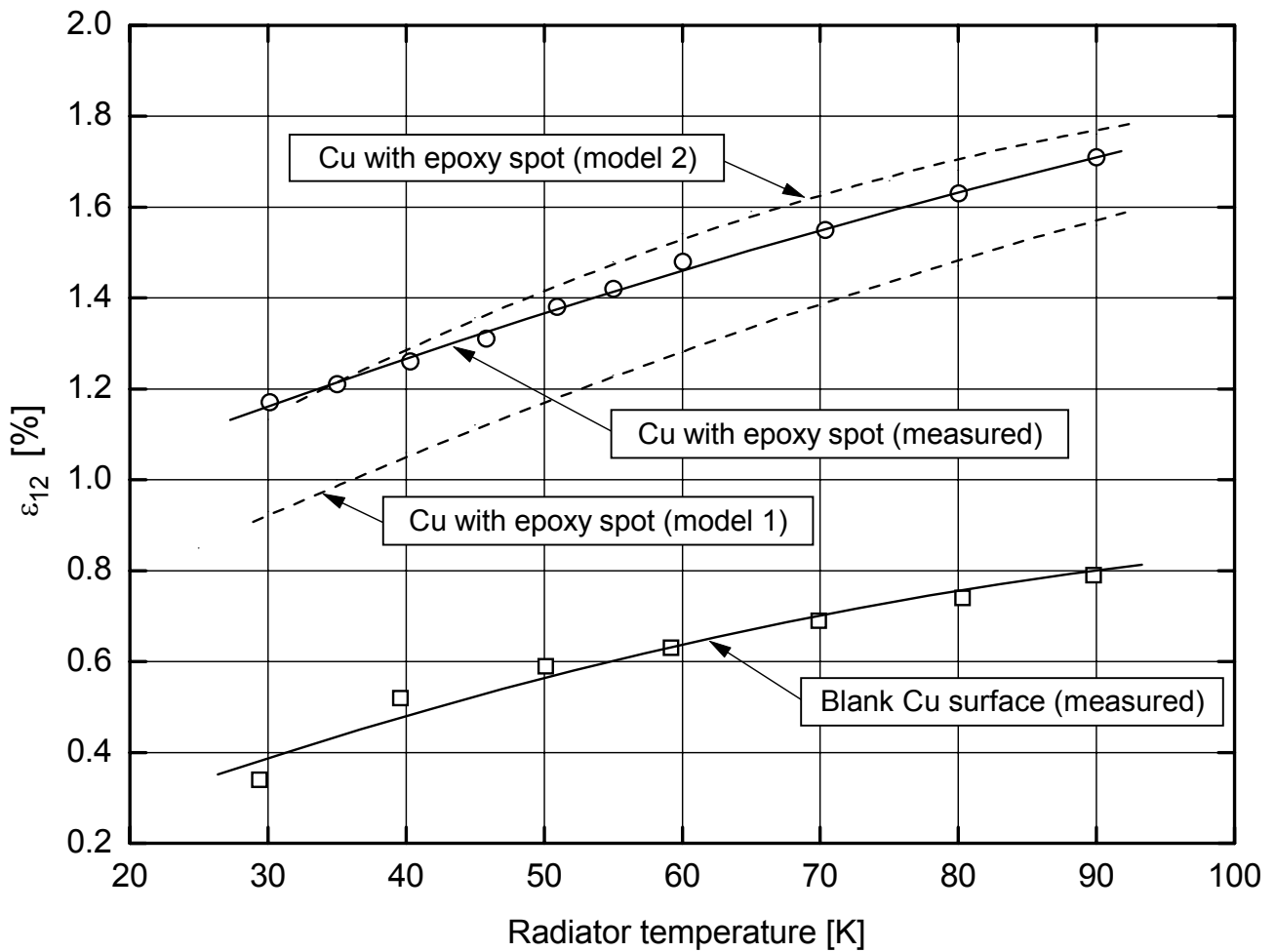

Figure 2 - Test of the method

To verify the correspondence of our results with published data, the measurement of Al (purity 99.5 %) at 5 K was compared with (Biondi and Guobadia, 1968, Bos and Lynch, 1970, Tsujimoto et. al., 1982, Ramanathan, 1952). Open symbols are used for optical (spectral) measurements and the filled ones for the thermal measurements plotted in Figure 3.

In the case of thermal measurements the temperature $T_R$ was converted to the wavelength by means of Wien's displacement law:

$$\lambda_{max} = \frac{3000}{T_R} \quad [\mu m; K] \qquad (3)$$

It can be proved for metals at low temperatures and far infrared radiation that the value of total hemispherical absorptivity is near to the value of spectral normal absorptivity measured at this wavelength.

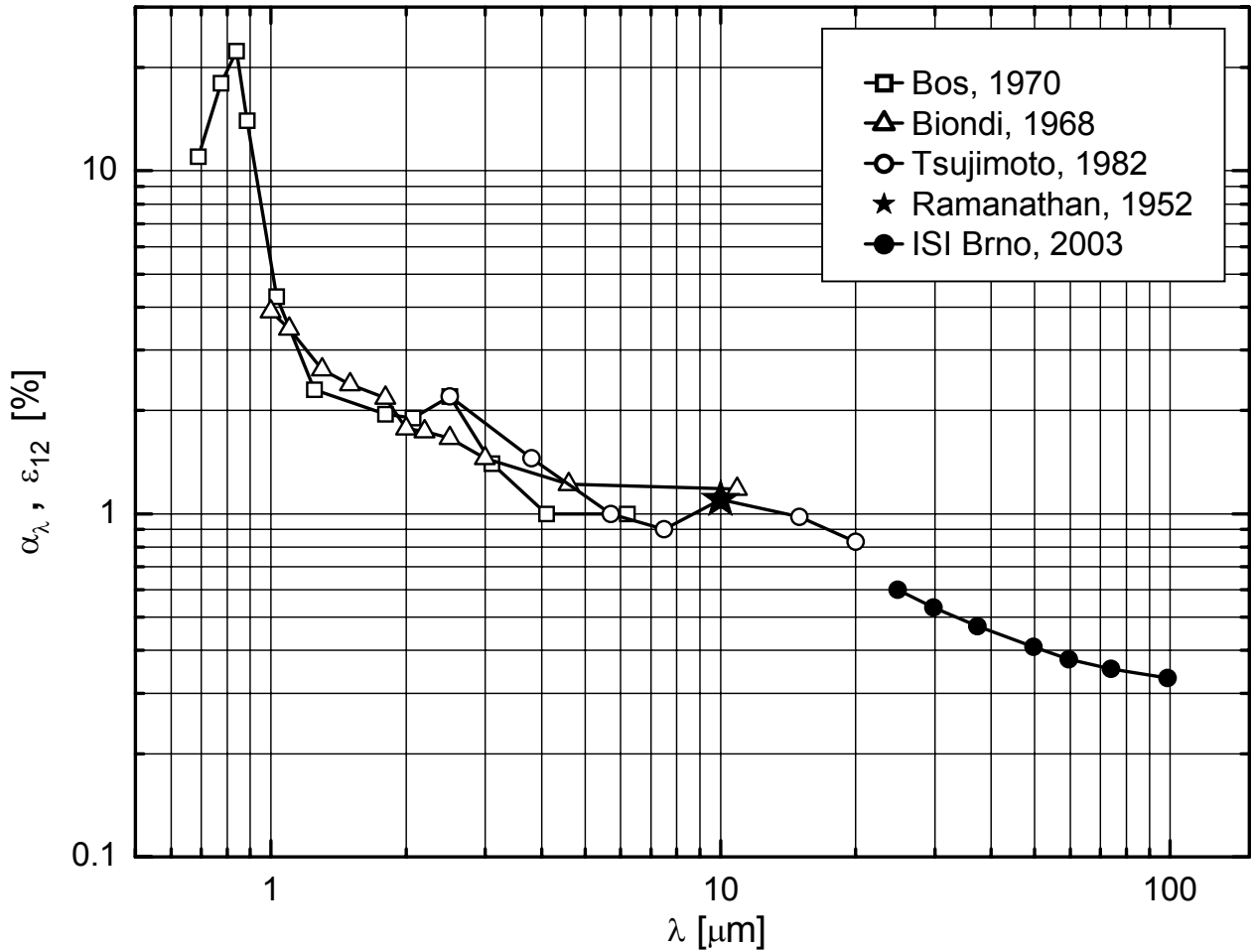

Figure 3 - absorptivity of aluminum at 4 K-5 K

### 5.2 EXAMPLES OF MEASUREMENTS

In Figure4 the experimental results of measuring of super insulation foils and an Al wrapping foil are shown. Foils were glued on a black epoxy surface. The firm Austrian Aerospace provided us with the first four foils listed in the legend of Figure 4. The samples Nr. 2151 and Nr. K 73 are 6 $\mu m$ polyester foils one side aluminized with a square resistance of 0.07-0.029 ohm/sq and double side aluminized with a square resistance of 0.5-0.7 ohm/sq., respectively. Nr. K 76 is a 12 $\mu m$ polyester foil, double side aluminized, with a square resistance of 0.5-0.8 ohm/sq. The 9 $\mu m$ aluminium foil of 99.5% purity, shiny surface with 0.01-0.02 ohm/sq., is labelled Nr. K 61. Furthermore, the absorptivity of the shiny and mat face of the wrapping foil was measured.

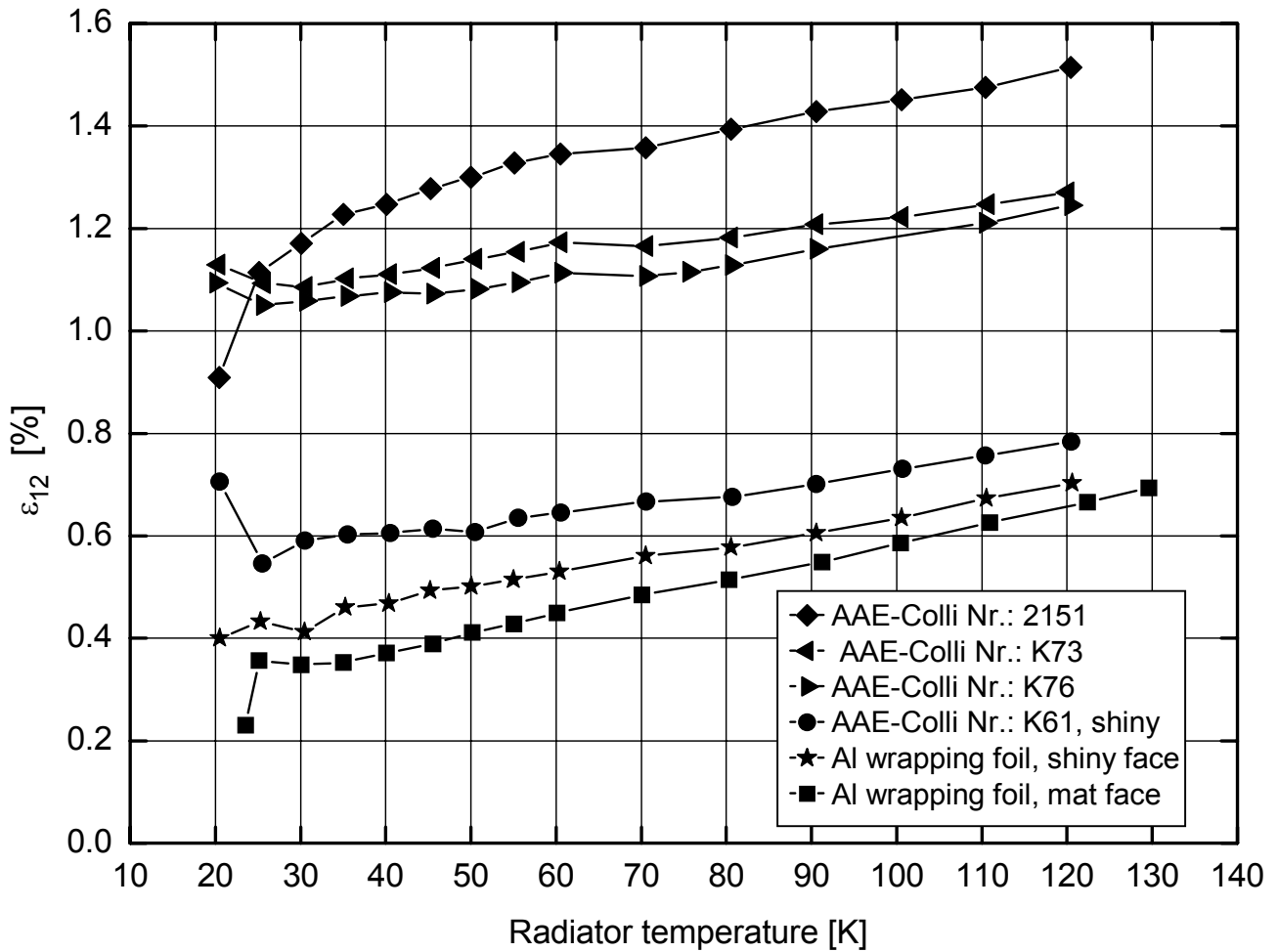

Figure 4 - emissivity of some superinsulation foils and Al-foils

## CONCLUSIONS

An apparatus for measurement of emissivity and absorptivity of material surfaces at cryogenic conditions was developed. In contradistinction to most of published data, temperature dependences of absorptivity were obtained at experiments. The device sensitivity enables the measurement of absorptivity above the value of 0.3 % when the sample is irradiated by a heat emitted from a black surface at 25 K. It is possible to use a commercial dewar can for the apparatus cooling. The advantage of the device is easy preparation of the samples. Measurement of one temperature dependence takes one day. Nowadays common cryogenic materials with various surface treatments are continuously measured (Musilova *et. al*., 2004). The measurements are very well reproducible. The application of this apparatus helps to extend the knowledge about radiative properties of materials and their temperature dependence.

## ACKNOWLEDGEMENT

This work is supported by the Academy of Sciences of the Czech Republic, the projects No. IBS2065109 and No. KSK 2067107.

We would like to thank the firm Austrian Aerospace GmbH for sending of samples of superinsulation foils.

# REFERENCES

1. Amano, T., Ohara, A., 1991, An experimental study of thermal radiation at cryogenic temperatures, *Heat Transfer-Japanese Research*, vol. 20, p. 307-323
2. Biondi M., A., Guobadia A. I., 1968, Infrared Absorption of Aluminum, Copper, Lead and Nickel at 4.2 K, *Phys. Rev.*, vol. 166, p. 667-673
3. Bos L. W., Lynch D. W. ,1970, Low energy optical absorption peak in aluminum and Al-Mg alloys, *Phys. Rev. Lett.*, vol. 25, p. 156-158
4. Gmelin, E., Asen-Palmer, M., Reuther, M., Willar, R., 1999, Thermal boundary resistance of mechanical contacts between solids at sub-ambient temperatures, *J. Phys. D: Appl. Phys.*, vol. 32, p. 19-43
5. Hanzelka, P., Musilova, V., 1995, Influence of changes in atmospheric pressure on evaporation rates of low-loss helium cryostats, *Cryogenics*, vol. 35, no 3, p. 215-218
6. Hanzelka, P., Jura, P., 2002, Fuzzy pressure controller for helium bath cryostats, *Proc.7$^{th}$ IIR Int. Conf. CRYOGENICS'2002*, Praha, p. 53-56.
7. Kaganer, M. G.,1969, *Thermal Insulation in Cryogenic Engineering*, IPST, Ltd., Jerusalem, 220 p.
8. Musilova, V., Hanzelka, P., Kralik, T., Srnka, A., 2004, Low Temperature Emissivity of Metals Used in Cryogenics, *Proc.8$^{th}$ IIR Int. Conf. CRYOGENICS 2004*, Praha, (*this Proceedings*)
9. Ramanathan, K.G., 1952, Infra-Red Absorption by Metals at Low Temperatures, *Proc. Phys. Soc. London A65*, p. 532-540
10. Scott, R. B., 1959, *Cryogenic Engineering*, New York, 368 p.
11. Tsujimoto, S., Kanda, M., Kunimoto, T., 1982, Thermal properties of some cryogenic materials, *Cryogenics*, vol. 22, p. 591-597